\documentclass[English,superscriptaddress,prl,twocolumn]{revtex4-1}
\usepackage{graphicx}
\usepackage{epstopdf}
\usepackage[ansinew]{inputenc}
\usepackage{array}
\usepackage{color}
\usepackage{amsmath}
\usepackage{amsxtra}
\usepackage{amstext}
\usepackage{amssymb}
\usepackage{latexsym}
\usepackage{float}
\usepackage{soul}
\usepackage[colorlinks=true, linkcolor=blue,urlcolor=blue,citecolor=blue]{hyperref}

\begin{document}

\title{Twistable electronics with dynamically rotatable heterostructures}
\author{Rebeca Ribeiro-Palau}
\thanks{R. R.-P. and Ch.Z. contributed equally to this work}
\altaffiliation[R.R.-P. Present address: ]{Centre de Nanosciences et de Nanotechnologies (C2N), CNRS, Univ Paris Sud,
Universit\'e Paris-Saclay, 91120 Palaiseau, France}
\email{rebeca.ribeiro@c2n.upsaclay.fr}
\affiliation{Department of Physics, Columbia University, New York, NY, USA}
\affiliation{Department of Mechanical Engineering, Columbia University, New York, NY, USA}
\author{Changjian Zhang$^{*}$}
\affiliation{Department of Mechanical Engineering, Columbia University, New York, NY, USA}
\affiliation{Department of Electrical Engineering, Columbia University, New York, NY, USA}
\author{Kenji Watanabe}
\affiliation{National Institute for Materials Science, 1-1 Namiki, Tsukuba, Japan}
\author{Takashi Taniguchi}
\affiliation{National Institute for Materials Science, 1-1 Namiki, Tsukuba, Japan}
\author{James Hone}
\affiliation{Department of Mechanical Engineering, Columbia University, New York, NY, USA}
\author{Cory R. Dean}
\affiliation{Department of Physics, Columbia University, New York, NY, USA}

\maketitle

{\bf The electronic properties of two-dimensional materials and their heterostructures can be dramatically altered by varying the relative angle between the layers.  This makes it theoretically possible to realize a new class of twistable electronics in which device properties can be manipulated on-demand by simply rotating the structure. 
Here, we demonstrate a new device architecture in which a layered heterostructure can be dynamically twisted, {\it in situ}.  We study graphene encapsulated by boron nitride where at small rotation angles the device characteristics are dominated by coupling to a large wavelength Moir\'e superlattice. The ability to investigate arbitrary rotation angle in a single device reveals new features in the optical, mechanical and electronic response in this system.  Our results establish the capability to fabricate twistable electronic devices with dynamically tunable properties.}

The weak van der Walls forces between the atomic planes in 2D materials makes it possible to fabricate devices with arbitrary rotational order. This provides a new opportunity in device design where electronic properties are controlled by varying the relative twist angle between layers \cite{Carr2017}.  Indeed several studies have established that in heterostructures assembled from 2D crystals, electron tunneling between layers varies strongly with rotation\cite{Britnell2013,Mishchenko2014,Greenaway2015,Fallahazad2015, Chari2016,Koren2016b,Wallbank2016}. In twisted bilayer graphene (two monolayers in direct contact but with an angle mismatch between the layers) several novel phenomenon have been predicted and observed, including topological valley transport \cite{Gonzalez-Arraga2017,Cao2016,Sanchez-Yamagishi2016,Ju2015,Kim2017}, and superconductivity\cite{Cao2018}, as a consequence of angle-dependent interlayer coupling. Likewise, the formation of interlayer excitons in transition metal dichalcogenide heterostructures is highly sensitive to angle\cite{Yu2015, Rivera2016,Rivera2015}. 

The effect of rotational alignment between conducting and insulating 2D layers can be equally significant.  A remarkable example is provided by graphene coupled to hexagonal boron nitride (BN).  Owing to the closely matched lattice constants a large Moir\'e superlattice develops near zero angle mismatch\cite{Yankowitz2012,Woods2014}.  This substantially alters the graphene band structure opening an energy gap at the charge neutrality point (CNP) and creating replica Dirac points at higher energies\cite{Hunt2013,Ponomarenko2013,Dean2013}.

Several techniques have been developed to fabricate layered heterostructures with controlled rotation between the layers, including optical alignment of crystal edges \cite{Yankowitz2012,Hunt2013,Ponomarenko2013,Dean2013}, rotational alignment \cite{Kim2016Matt} during assembly, and self-alignment through thermal annealing \cite{Wang2016, Woods2016}). However, in each case \textit{a priori} understanding of the crystallographic orientation of each layer is required before assembly;  motion between the layers during assembly makes it difficult to achieve precise angle control; and most significantly,  once assembled the angle can not be further modified. Here, we present a new experimental technique that provides on-demand control of the orientation between layers in a van der Waals heterostructure. We study a BN/graphene/BN structure where we demonstrate {\it in situ} control over the length of the Moir\'e potential and consequently achieve dynamic tunability of the optical, mechanical and electrical properties of the system.

\begin{figure*}[t!]
\includegraphics[scale=0.6]{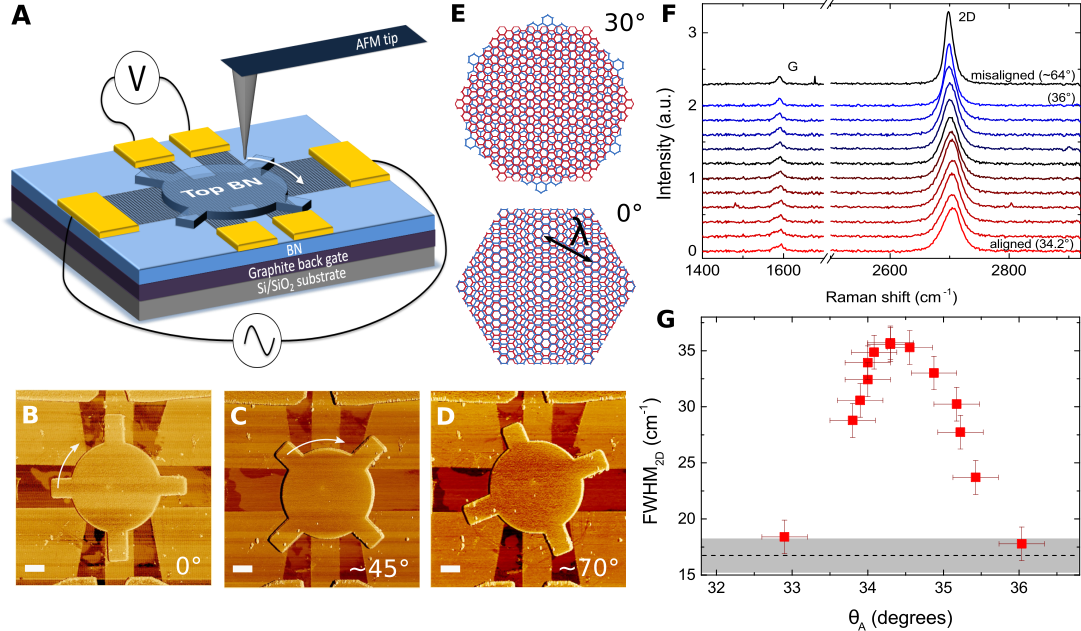}  
\caption{{\bf Rotatable heterostructures} {\bf a}, Schematic cartoon of the device structure and the experimental technique. {\bf b-d}, AFM image of a fabricated device showing three different orientations of the top BN. The angles identified in each panel is the absolute angle referenced to the AFM coordinate system (labelled $\theta_A$ in the text). The images were acquired by the same AFM used to rotate the BN layer. {\bf e} Schematic illustration of the Moire superlattice arising between graphene (Red) and BN (Blue) at zero angle. The  moire wavelength is identified by $\lambda$. {\bf f}, Raman spectrum of the device shown in (b-c) for $\theta_A$ between 34.2$\pm$0.2 degrees and 36$\pm 0.2$.  Black curve shows an additional measurement acquired at  $\approx64$ degrees. {\bf g}, FWHM of the 2D peak as a function of the absolute angle. All Raman measurements were taken with the gate bias held at $V_{G}=0$~V. The peak FWHM position identifies  zero angle crystallographic alignment (see text). Dashed line represents the FWHM measured for all angles larger than approximately 2 degrees away from perfect alignment with the shaded region representing the associated uncertainty.}
\label{Device-AFM}
\end{figure*}

Figure \ref{Device-AFM}a  shows a cartoon schematic of our device design.  Using the mechanical assembly technique\cite{Wang2013}, graphene is placed on top of a large flake of BN, and then etched into a Hall bar shape using oxygen plasma. The graphene layer is intentionally misaligned to this BN, producing a short-wavelength Moir\'e potential that does not significantly alter the intrinsic graphene band structure \cite{Dean2010}. Next, a pre-shaped BN structure is transferred on top of the graphene. Finally, electrical contacts are patterned onto the exposed leads of the graphene (see supplementary information). Due to the low mechanical friction between graphene and BN, we are able to freely rotate and translate this top BN layer using an atomic force microscope (AFM). Pushing on one of the arms of the uppermost BN structure rotates this layer (Fig. \ref{Device-AFM}b-d) changing its crystallographic orientation with respect to the graphene layer.  The Moir\'e superlattice wavelength, $\lambda$, generated between these crystals is given by:

\begin{equation}
\lambda= \frac{(1+\delta)a}{\sqrt{2(1+\delta)(1-cos(\theta))+\delta^2}}
\label{angle}
\end{equation}

\noindent where $\delta=0.017$ is the lattice mismatch between graphene and BN, $a$ is the lattice constant of graphene  and $\theta$ is the rotational mismatch between the layers.  By rotating the top BN layer, the wavelength of the resulting Moir\'e superlattice can therefore be dynamically varied. 

\begin{figure*}[t!]
\includegraphics[scale=0.5]{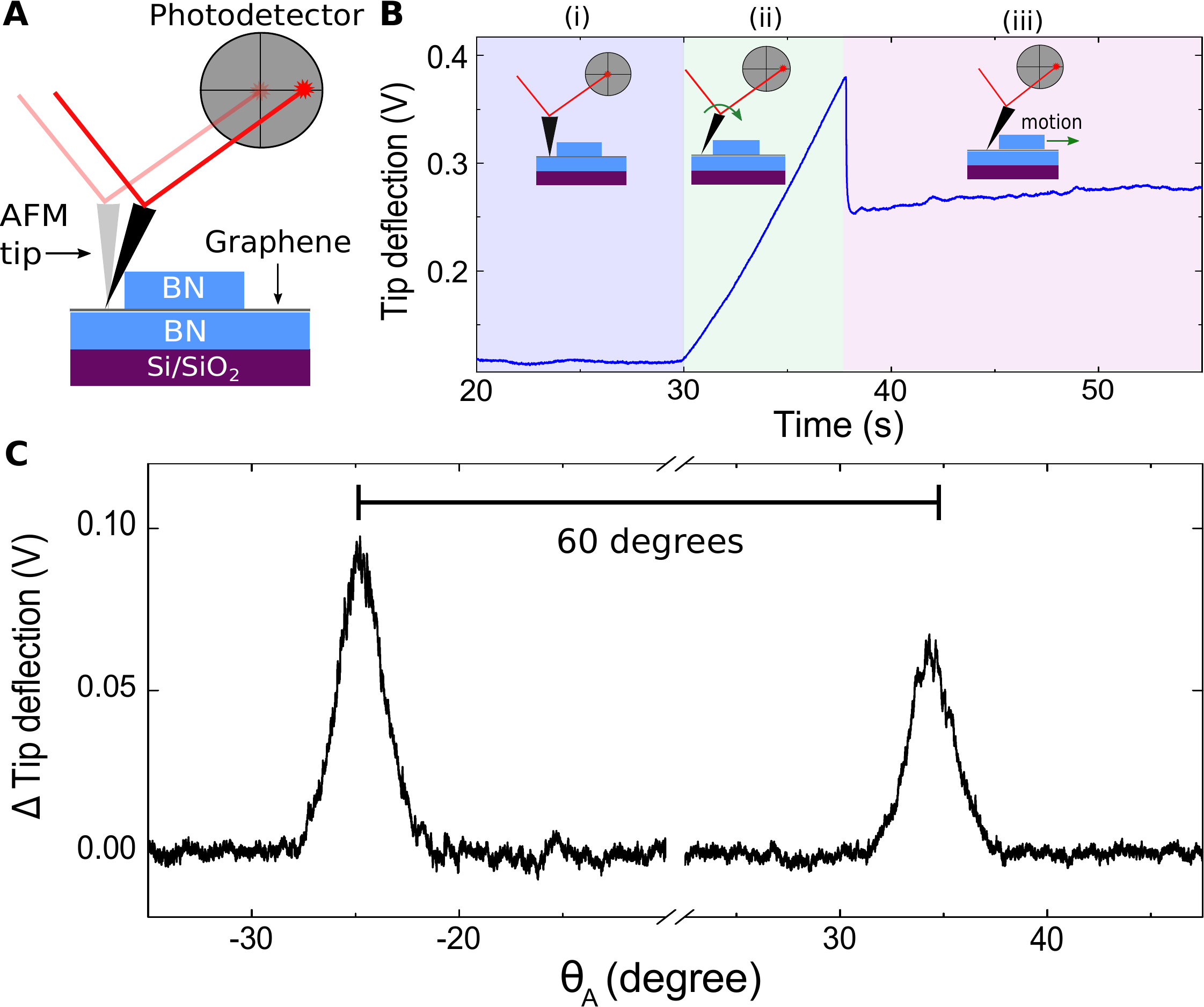}  
\caption{{\bf Mechanical properties versus angle.} {\bf a}, Schematic description of friction measurements.  When the AFM tip encounters the BN structure it cants causing a repositioning of the reflected laser spot in the four-quadrant photodetector.  The resulting voltage difference is proportional to the tip cant angle (referred to here as the tip deflection) and serves as a measure of the torque force acting at the end of the tip. {\bf b}, Tip deflection versus time in a translational push of the  upper BN structure. Different regimes of the measurement are identifiable: (i) as the tip drags along the surface, tip-substrate friction results in a steady state tip deflection (ii) when the tip encounters the BN structure it initially resists translation and the tip deflection increases. We refer to this as the static friction regime (iii) Once the BN is in motion the tip deflection relaxes slightly providing a measure of the dynamic friction at the BN-graphene interface. {\bf c}, tip deflection versus absolute angle measured during a continuous rotation of the BN. Two peaks are observed, spaced 60 degrees apart.}
\label{Friction}
\end{figure*}

Imaging with the same AFM provides a real-time measurement of the orientation of the rotatable BN layer relative to the reference frame of the AFM, which we label by the {\it absolute angle} ($\theta_{\rm A}$) (see for example Fig. \ref{Device-AFM}b-d).  More significant is the {\it relative angle}, $\theta$, between the rotatable BN and encapsulated graphene crystal lattices.  To determine this we identify and measure angle-dependent features in the optical Raman spectrum of the heterostructure.  

Figure \ref{Device-AFM}f shows a series of Raman spectra measured from $\theta_{\rm A}=34-64$ degrees. The most striking variation in the Raman spectra is an increase in the full width at half maximum of the 2D peak  (FWHM$_{\rm 2D}$) \cite{Eckmann2013,Woods2014}.  Plotting this as a function of absolute angle (Fig. \ref{Device-AFM}g) we see a well defined maximum, occurring for this device at $\theta_{\rm A}\sim34.2$ degrees.  Based on previous work \cite{Eckmann2013}, we interpret the peak position as corresponding to zero angle alignment between the BN and graphene crystals. This therefore provides a reference which we can use to determine $\theta$ for any orientation of the BN layer.  

 Our result highlights the robustness of Raman spectroscopy as a tool to characterize the rotational order in these van der Waals heterostructures.  Moreover, we emphasize that previous efforts to study effects of twist angle in this system required numerous samples with different fixed angles whereas here we demonstrate a mapping of the angular dependence with better than 0.2 degree precision, in a single tunable device.

One notable disagreement with previous results\cite{Eckmann2013} is an overall narrower 2D line-width in our devices. In previous work this reduced linewidth was proposed to be a consequence of a reduction of the in-plane strain in fully-encapsulated structures, such that areas of full commensuration to the aligned BN disappear \cite{Woods2014}. However, our observation of the same linear trend and a reduced linewidth for the misaligned position - for which no commensurate regions are expected - suggests a different origin.  We speculate that the real origin of this linewidth reduction is the change of dielectric environment of the graphene by the presence of the second BN, as proposed in ref. \cite{Neumann2015}. Therefore, our results re-open the question of the existence of commensurate states in encapsulated graphene devices.  We additionally have identified a shift in the position of the 2D and G peaks with relative angle (see supplementary information), a full discussion of which is beyond the scope of this report.

\begin{figure*}[t!]
\includegraphics[scale=0.5]{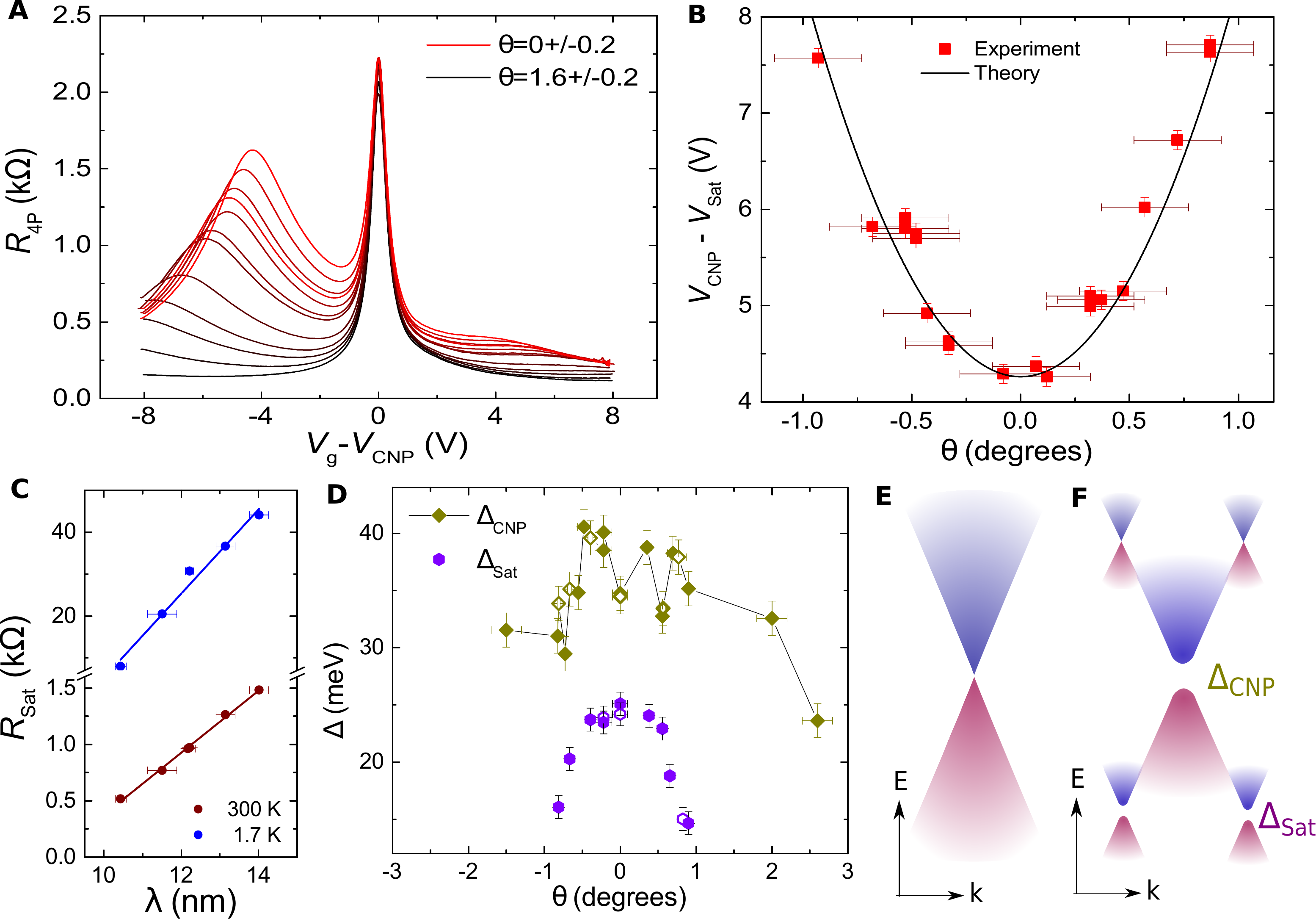}  
\caption{{\bf Electronic transport properties.} {\bf a}, Four-terminal resistance as a function of the gate voltage for different alignments of the graphene/BN structure, acquired at room temperature.  {\bf b}, Position of the satellite peak in gate voltage as a function of the relative  angle. The 0.2 degrees error bar in the angle reports the precision achieved with the AFM imaging in tapping mode.  {\bf c}, Linear dependence of the maximum value of the four-terminal resistance at the satellite peak at 300 K  (red) and 1.7 K (blue) as a function of the Moir\'e length. {\bf d} Energy gap, measured by thermal activation, for the satellite peak (circles) and the charge neutrality point (diamonds) as a function of the relative angle. Open symbols represent a repeated measurement at a given angle after moving through other angles and thermally cycling. {\bf e}, Schematic band structure for native graphene and {\bf f}  for a graphene-hBN heterostructure with a small twist angle.}
\label{Transport}
\end{figure*}

Mechanical resistance while pushing the BN imparts a torque to the AFM tip, causing it to cant away from a vertical position and produce a voltage difference in the AFM's photodetector. This can be  used to identify variations in frictional forces (Fig. \ref{Friction}a.) When sliding the top BN layer with the AFM tip, in a translational motion far from alignment, we identify three regimes (see Fig. \ref{Friction}b): i) sliding friction between the tip and the substrate before the tip encounters the BN; ii) static friction when the tip encounters the BN but it resists translation; and iii) dynamic friction once the BN begins to move.

Figure \ref{Friction}c shows a plot of the  change in the tip deflection under continuous rotation (dynamic friction), where the background due to piezoelectric drift and residual friction has been subtracted.  Two prominent peaks appear, separated by 60 degrees. This closely resembles previous measurement of friction between two graphitic structures in which a transition from superlubricity (where the structures are in an incommensurate position and the atomic shear forces are negligible) to a dissipative state was observed at commensurate angles of the 3-fold symmetric hexagonal lattices\cite{Dienwiebel2004,Liu2012,Filippov2008}.  However, since there is a lattice mismatch between graphene and BN there is not true lattice commensurability at any angle and therefore the increase of the friction should have a different origin. Recent numerical simulations suggest that contributions of the Moir\'e superlattice to the frictional force cannot be neglected and are expected to be maximal for aligned layers \cite{Koren2016}, which could explain our experimental result. A more detailed study of the interlayer frictional forces will be necessary to fully understand this behavior.  While this is beyond the present scope, we note that this new device structure allow us to study mechanical properties, such as frictional force, in atomically flat materials without rugosity contributions. These results also highlight the possibility to use the friction response as an {\it in situ} method to monitor and control layer alignment in heterostructures.

\begin{figure}[t!]
\includegraphics[scale=0.75]{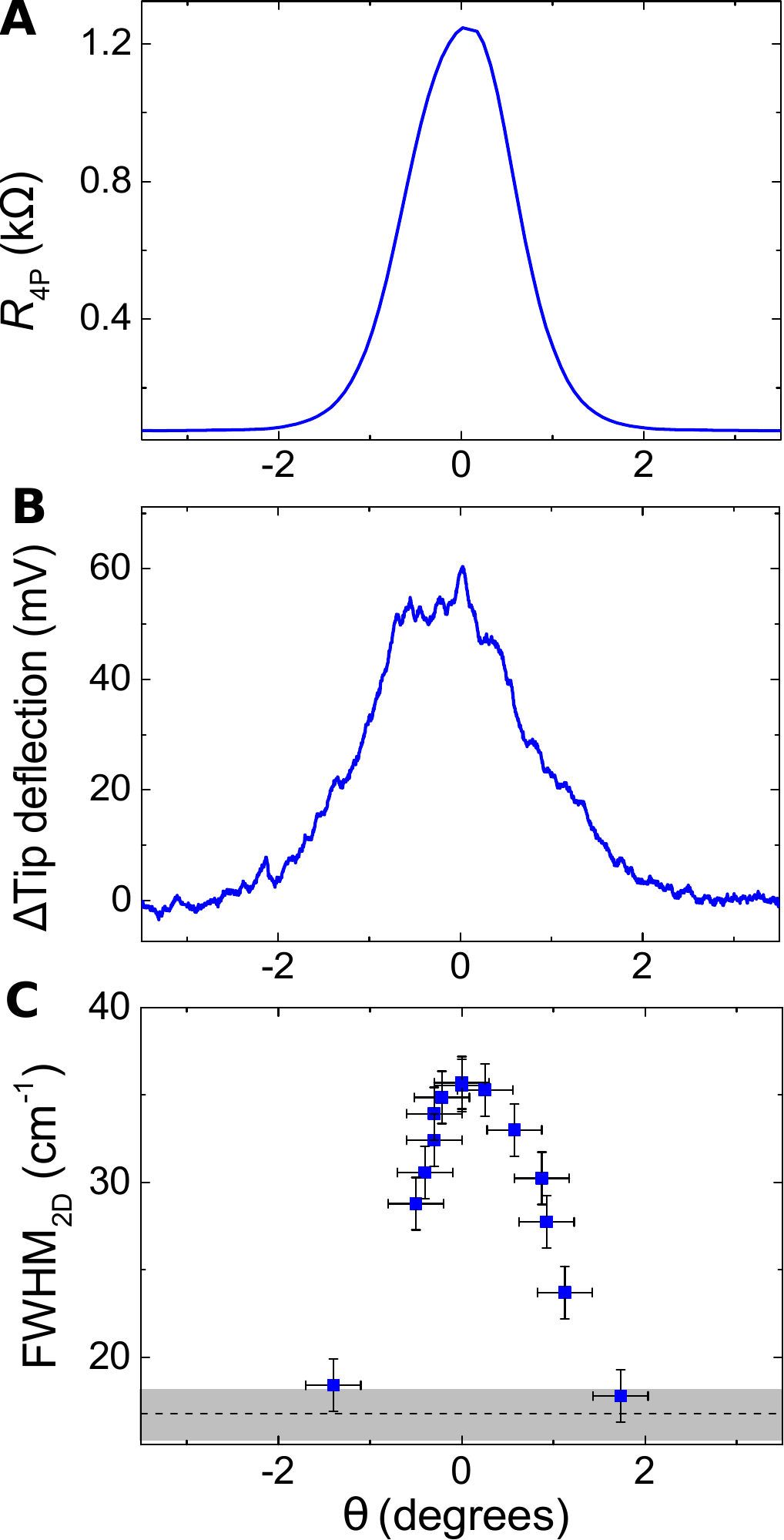}  
\caption{{\bf Compilation of the angle control technique.}  {\bf a} Four-terminal resistance as a function of the relative angle measured at a carrier density of -1.9x10$^{12}$ cm$^{-1}$. {\bf b}, Tip deflection in friction measured simultaneously as the electronic transport.  {\bf c}, FWHM of the 2D peak of the Raman spectrum as a function of the angle. All measurements were performed in the same device.}
\label{Merit}
\end{figure}

Our device design allows us to measure electron transport in the active layer (in this case graphene) while changing the relative orientation of the over-layer. Figure \ref{Transport}a shows a plot of the four-terminal resistance of the graphene layer as a function of back gate voltage $V_g$ for different values of $\theta$ at room temperature. Near $\theta=0$, additional satellite resistance peaks appear symmetrically in density around the charge neutrality point (CNP).  This is consistent with the emergence of satellite Dirac points induced by scattering from the Moir\'e superlattice potential \cite{Hunt2013,Ponomarenko2013,Dean2013,Yankowitz2012}. 

As $\theta$ increases away from zero, the satellite peaks diminish in intensity and moves further from the CNP to higher gate values. To analyze this behavior more quantitatively, Figure \ref{Transport}b plots the satellite peak position in $V_g$ versus $\theta$ determined from the AFM imaging. The measured position shows excellent agreement with the values of carrier density at which the full filling of the miniband occurs, $n=8/\sqrt{3}\lambda^2$, where the carrier density and gate voltage are related by $n=C_{\rm g}(V_{\rm g}-V_{\rm CNP})$ and $\lambda$ is given by expression (\ref{angle}) \cite{Hunt2013,Ponomarenko2013,Dean2013,Yankowitz2012}. The error bars in Fig. \ref{Transport}b,  approximately $\pm 0.2$ degrees, reflects the precision with which $\theta$ can be determined from the AFM topographic images. Determining $\theta$ from the gate voltage position of the satellite peak at low temperature provides a more accurate measurement with uncertainty less than $<\pm0.1$ degrees, however this method of determining the angle is limited to the only a few degrees where the satellite peak remains within an accessible density range.

Plotting the resistance of the satellite peak versus $\lambda$ for a single device reveals an apparently linear variation in the magnitude of the satellite peak resistance (Fig. \ref{Transport}c). Interestingly, the resistance of the CNP does not change linearly with $\theta$ in this range (see supplementary information). The origin of this linear dependence is not known at present.  The striking observation however highlights an example of a physical phenomenon previously obscured\cite{Hunt2013,Ponomarenko2013,Dean2013,Yankowitz2012} by sample-to-sample variations but which becomes clear when able to measure the effect of varying rotation in single sample. 

We measured the energy gap versus angle at both the central and satellite Dirac points by thermal activation.  The gap magnitudes, shown in Fig. \ref{Transport}d, are in good agreement with electronic transport  measurements in encapsulated \cite{WangLei2015} and non-encapsulated devices \cite{Hunt2013,Woods2014}, optical measurements made in epitaxial BN/graphene structures \cite{Chen2014}, and theoretical calculations \cite{Song2013}. As shown in Figure \ref{Transport}d, the energy gap of the satellite peak decreases smoothly away from $\theta=0$. In contrast, the energy gap at the CNP displays a more complex behavior near $\theta=0$ and only decreases significantly for $|\theta|>2$ degrees.  The origin of the extra structure observed for the CNP remains to be understood. However, this does not appear to be simple experimental noise since several gap values were confirmed to be reproducible when measured non-consecutively ({\it i.e,} after thermally cycling and rotating through  different angles and back). The difference in behavior of the two energy gaps, impossible to observe in study utilizing multiple samples at fixed angles \cite{WangLei2015}, reflects their different physical origins and highlights the importance of our new technique in the fully understanding the band structure modifications resulting from variations in angular alignment, Fig. \ref{Transport}e-f. The persistance of an energy gap at the CNP in encapsulated devices for angles beyond the angle  at which a commensurate-incommensurate transition was previously identified \cite{Woods2014} suggests that this energy gap is not related solely to the presence of a commensurate state.

Fig. \ref{Merit} shows a direct comparison of the optical, mechanical and  electrical response versus angle measured in the same device. The four probe resistance (Fig. \ref{Merit}a), the maximum tip deflection signal (friction, Fig. \ref{Merit}b) and the maximum FWHM of the Raman 2D  peak (Fig. \ref{Merit}c) coincides exactly, confirming the relationship between these properties. The  four-probe resistance (Fig. \ref{Merit}a) and friction response (Fig. \ref{Merit}b) were acquired simultaneously at a fixed carrier density of 1.9x10$^{12}$ cm$^{-1}$ (corresponding to the dashed line in Fig. S2) while continuously rotating the BN layer. We  note that at this relatively large carrier density the bulk resistance is modulated by more than an order of magnitude over less than 2 degrees of rotation (at cryogenic temperatures this increases is of more than two orders of magnitude, see supplementary information). Our demonstration that rotatable heterostructures with dynamically tunable device characteristics can be realized provides a new opportunity in device engineering.  While here we have investigated BN encapsulated graphene as a model tunable system, this technique is readily extended to generic heterostructures fabricated from 2D materials where in addition to band structure tunability, emergent phases such as superconductivity and magnetism may be  controllably varied with rotation.

\bibliography{references}

\section*{Acknowledgements}

 We acknowledge discussions with Matthew Yankowitz and Andres Botello-Mendez as well as Juan Huerta, Shaowen Chen and Martin Gustafsson for technical support. This research was supported by the NSF MRSEC programme through Columbia in the Center for Precision Assembly of Superstratic and Superatomic Solids (DMR-1420634). C.R.D. acknowledges partial support by the National Science Foundation (DMR-1462383).
 
 \section*{Author Contributions }
 
R.R.-P. and C.R.D. designed the experiment. Ch.Z. and R.R.-P. fabricated the samples, performed the experiments, analyzed the data and wrote the paper. T.T. and K.W. grew the crystals of hexagonal boron nitride. J.H. and C.R.D. advised on experiments, data analysis and writing the paper.





\renewcommand{\thefigure}{S\arabic{figure}}
\renewcommand{\thesubsection}{S\arabic{subsection}}
\renewcommand{\theequation}{S\arabic{equation}}
\setcounter{figure}{0} 
\setcounter{equation}{0}


\section*{Supplementary Information}

\begin{figure*}
\centering
\includegraphics[scale=0.5]{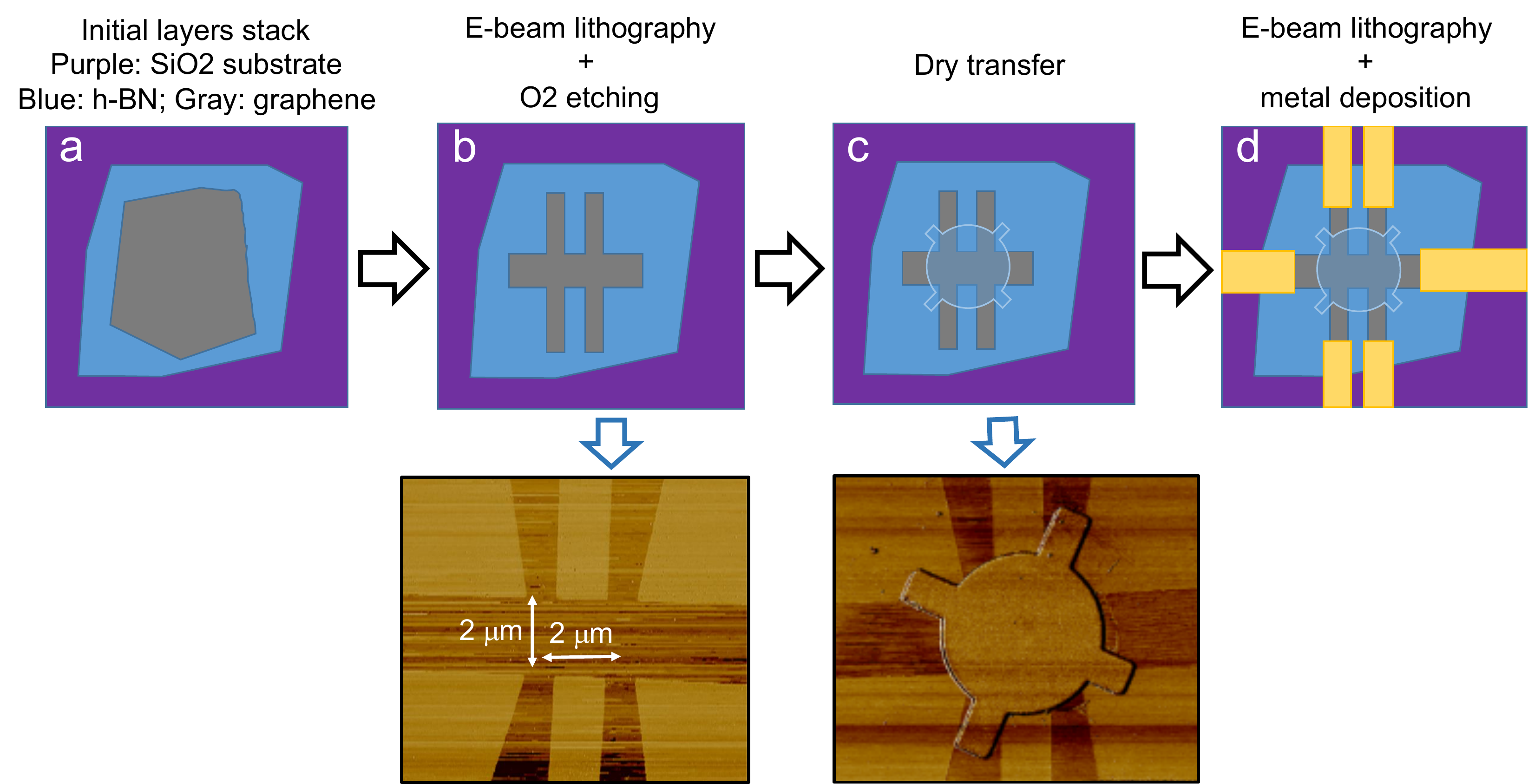}  
\caption{{\bf Fabrication process}. {\bf a}, Initial graphene/BN stack deposited on a Si/SiO$_2$ substrate {\bf b}, etching of the graphene flake to give a Hall bar shape. {\bf c}, Dry transfer of a pre-shaped BN structure. {\bf d}, electrical connection with Cr/Pd/Au top surface contacts.}
\label{Fabrication}
\end{figure*}

\section*{Raman spectrum of aligned structures}

The inset in Fig. \ref{Raman}a shows a plot of the FWHM of the 2D peak versus calculated Moir\'e wavelegth.   A remarkably linear dependence, described by the linear fit FWHM$_{\rm 2D}=2.7\lambda-0.77$,  is observed. The slope of this linear fit   is identical, within experimental uncertainty, to previous observations \cite{Eckmann2013}.  Previous theoretical studies of the modification of the Raman spectra by a superlattice potential associated this changes with the folding of the phonon structure due to the Moir\'e pattern\cite{Lockwood1987,Sood1985}, which could also be the case in graphene. 

\begin{figure*}
\centering
\includegraphics[scale=0.55]{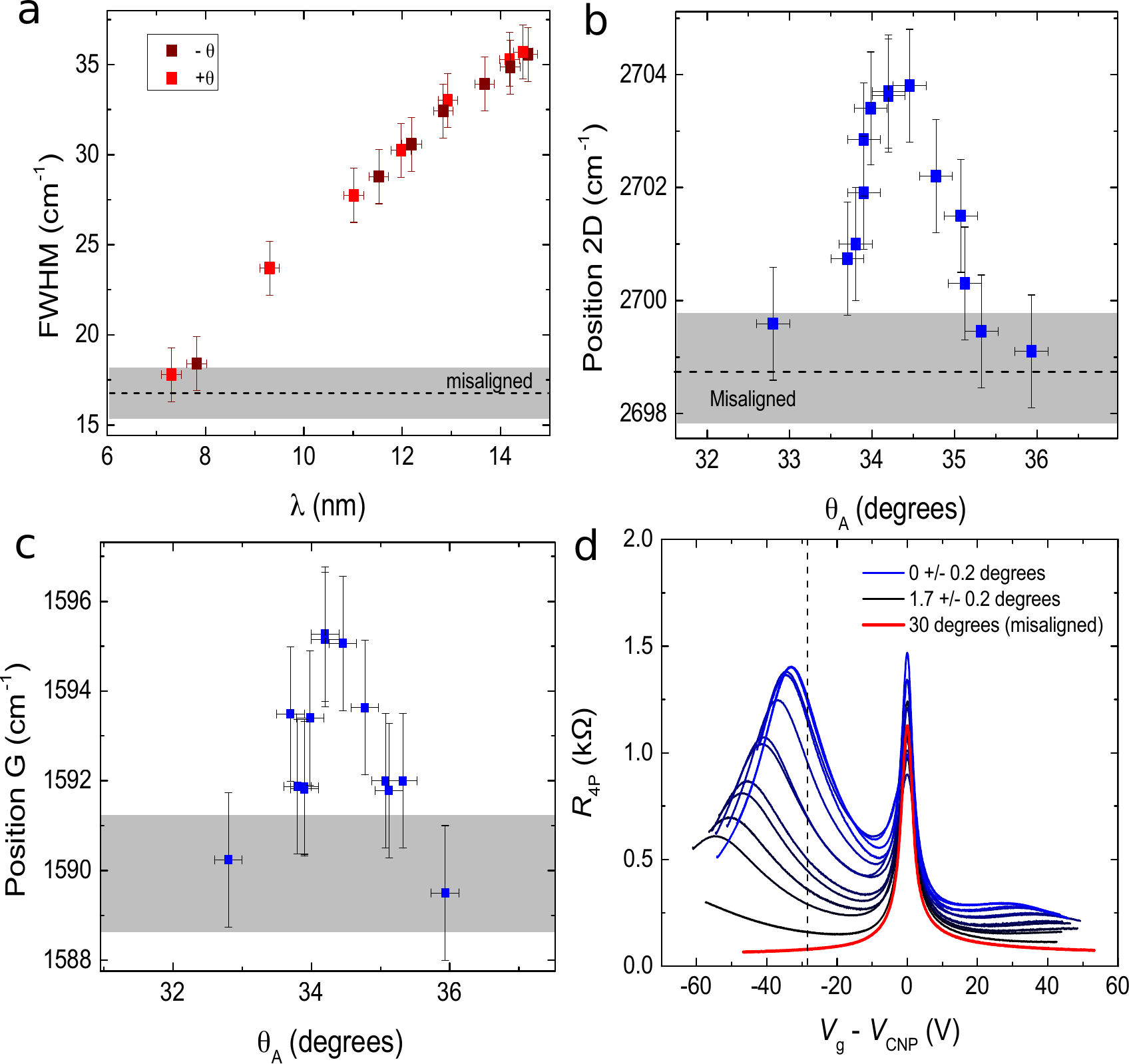}  
\caption{{\bf Raman spectrum}. {\bf a}, FWHM of the 2D peak as a function of the Moir\'e length, notice that the slope of the curve is the same reported in \cite{Eckmann2013}, see main text. Position of the 2D {\bf c}  and G {\bf d} peaks as a function of the absolute angle. {\bf d}, Room temperature measurement of the four-terminal resistance as a function of gate voltage for each Raman spectrum of this figure and of figure 1 of the main text. {\bf e}, Maximum of the resistance at the satellite peak as a function of the Moir\'e wavelength and energy calculated using $n=8/\sqrt{3}\lambda^2$ and $E=hv_{\rm F}/\sqrt{3}\lambda$, where the capacitive coupling of $C_{\rm g}/e=6.72\times10^{14}$ V$^{-1}$ m$^{-2}$ was obtained experimentally from Hall measurements.}
\label{Raman} 
\end{figure*}

\section*{Angle and Moir\'e length relation}

The angle between layers, $\theta$, and the Moir\'e length, $\lambda$ are related by

\begin{equation}
\lambda= \frac{(1+\delta)a}{\sqrt{2(1+\delta)(1-cos(\theta))+\delta^2}}
\label{angle}
\end{equation}

where $\delta=0.017$ is the lattice mismatch, $a$ is the lattice constant of graphene \cite{Yankowitz2012}. As explained in the main text the energy of this satellite peaks is given by:

 \begin{equation}
E=\pm\frac{h v_{F}}{\sqrt{3}\lambda} 
\label{energy}
\end{equation}

 As previously reported in \cite{Yankowitz2012} we found that the satellite peak is weaker for the conduction band (positive V$_g$) than the one of the valence band (negative V$_g$). This asymmetry in strength has been associated with the breaking of the electron-hole symmetry induced by a modulated hopping between different graphene sub-lattices.

\begin{figure*}
\centering
\includegraphics[scale=0.55]{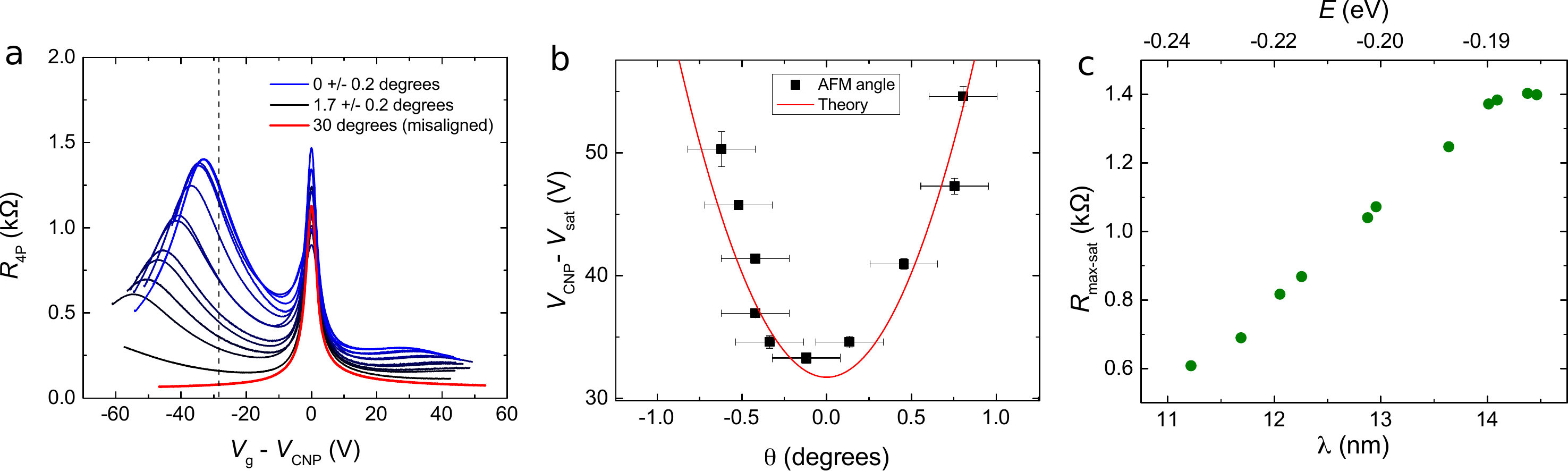}  
\caption{{\bf Satellite peaks}. {\bf a}, four probes resistance as a function of the back gate for different angles. Dashed lines represents the carrier density used when measuring simultaneaously resistance and friction in Figure 4 of main text. {\bf b} position of the satellite peak in gate voltage as a function of the  angle measured witht he AFM (where the minumum has been reported as zero. Solid line represents numerical fit using eqs \ref{angle} and \ref{energy}.  {\bf d} Resistance of the satellite peak as a functon of the Moir\'e length. }
\label{Raman}
\end{figure*}

\begin{figure*}
\centering
\includegraphics[scale=0.25]{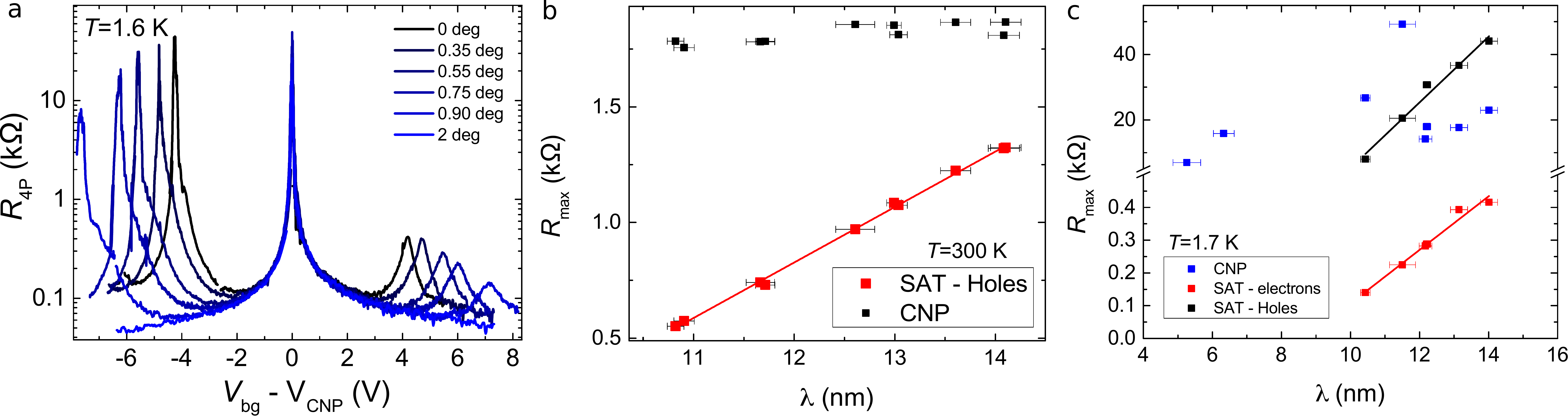}  
\caption{{\bf Linear resistance dependence}. Value of the four probe resistance at the charge neutrality point and satellite peaks at  300 K {\bf a} and 1.6 K {\bf b} for a second sample. The linear fits for the second sample can be described by $R_{\rm max-h(300K)}= 275.26\lambda-2376$, $R_{\rm max-e(1.6 K)}= 81.8\lambda-711$ and $R_{\rm max-h(1.6 K)}= 10040\lambda-95141.9$}
\label{Raman}
\end{figure*}

\begin{figure*}
\centering
\includegraphics[scale=0.25]{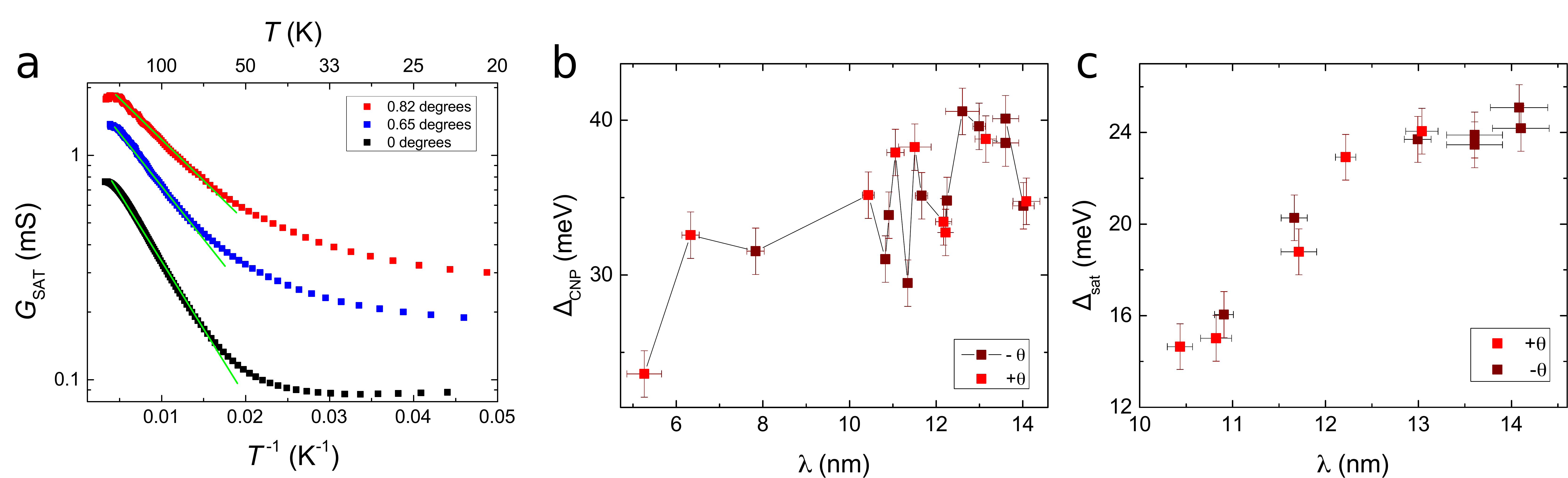}  
\caption{{\bf Energy gaps}. {\bf a}, Arhenious plot for satellite peak at three different angles.  Energy gap of the CNP  {\bf b} and satellite peak {\bf c} as a function of the Moir\'e length for the measurements reported in the main text.}
\label{Raman}
\end{figure*}

\begin{figure*}
\centering
\includegraphics[scale=0.4]{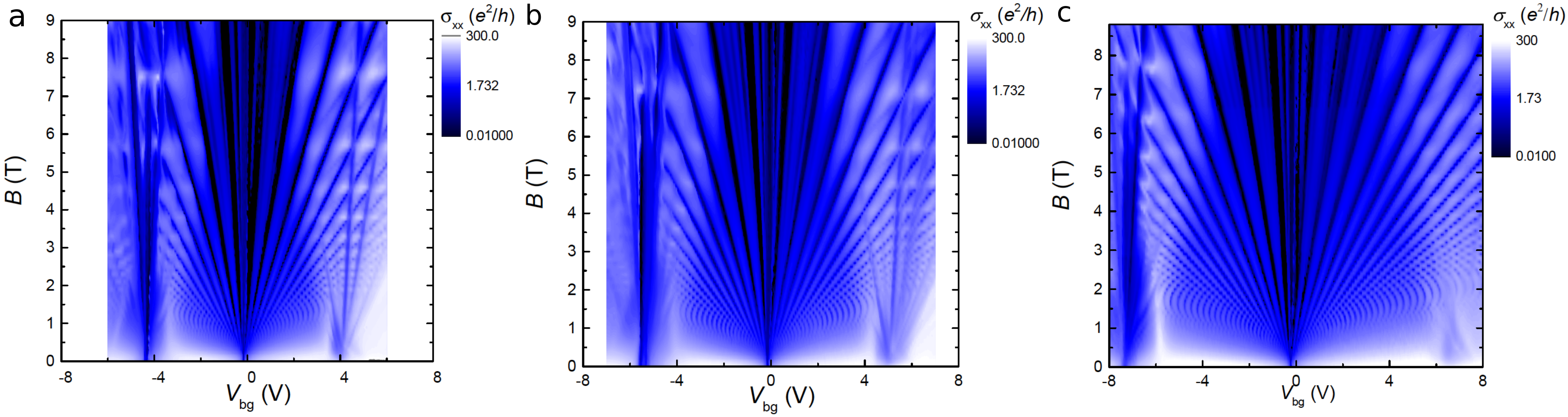}  
\caption{{\bf Magneto transport}. Longitudinal conductivity as a function of back gate voltage and magnetic field for 0 degrees ($\lambda=14.4\pm0.2$ nm) {\bf a}, -0.47 degrees {\bf b} and 0.83 degrees {\bf c}. Moir\'e length is calculated from the fits to Landau level crossing of Hofstadter spectrum and Hofstadter oscillations \cite{Hunt2013}.}
\label{Raman}
\end{figure*}

\begin{figure*}
\centering
\includegraphics[scale=1]{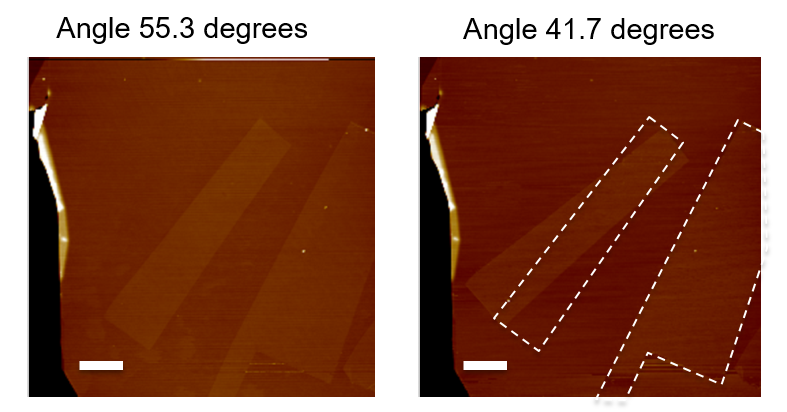}  
\caption{{\bf Rotation of monolayer graphene}. {\bf Left}, Graphene on BN as fabricated. {\bf Rigth} same graphene/BN stack after rotation of one monolayer. Dashed lines are guides to the eye with the shape and orientation of the left image. Scale bar 3 $\mu$m.}
\label{Raman}
\end{figure*}

\end{document}